\documentclass[preprint]{aastex63}

\usepackage{color}

\received{}
\revised{}
\accepted{}
\submitjournal{ApJL}

\shorttitle{ALMA view of the $\rho$ Ophiuchi A PDR with a 360-au beam}
\shortauthors{Yamagishi et al.}

\graphicspath{{./}{figures/}}

\begin{document}

\title{ALMA view of the $\rho$ Ophiuchi A PDR with a 360-au beam: the [C$\;${\sc i}] emission originates from the plane-parallel PDR and extended gas}

\correspondingauthor{Mitsuyoshi Yamagishi}
\email{yamagishi@ioa.s.u-tokyo.ac.jp}

\author[0000-0002-6385-8093]{Mitsuyoshi Yamagishi}
\affiliation{Institute of Astronomy, The University of Tokyo, 2-21-1 Osawa, Mitaka, Tokyo 181-0015, Japan}
\affiliation{National Astronomical Observatory of Japan, National Institutes of Natural Sciences, 2-21-1 Osawa, Mitaka, Tokyo 181-8588, Japan}

\author[0000-0001-9368-3143]{Yoshito Shimajiri}
\affiliation{National Astronomical Observatory of Japan, National Institutes of Natural Sciences, 2-21-1 Osawa, Mitaka, Tokyo 181-8588, Japan}

\author[0000-0002-2062-1600]{Kazuki Tokuda}
\affiliation{Graduate School of Science, Osaka Prefecture University, 1-1 Gakuen-cho, Naka-ku, Sakai, Osaka 599-8531, Japan}
\affiliation{National Astronomical Observatory of Japan, National Institutes of Natural Sciences, 2-21-1 Osawa, Mitaka, Tokyo 181-8588, Japan}

\author[0000-0002-8049-7525]{Ryohei Kawabe}
\affiliation{National Astronomical Observatory of Japan, National Institutes of Natural Sciences, 2-21-1 Osawa, Mitaka, Tokyo 181-8588, Japan}
\affiliation{The Graduate University for Advanced Studies (SOKENDAI), 2-21-1 Osawa, Mitaka, Tokyo 181-0015, Japan}

\author[0000-0001-5431-2294]{Fumitaka Nakamura}
\affiliation{National Astronomical Observatory of Japan, National Institutes of Natural Sciences, 2-21-1 Osawa, Mitaka, Tokyo 181-8588, Japan}
\affiliation{The Graduate University for Advanced Studies (SOKENDAI), 2-21-1 Osawa, Mitaka, Tokyo 181-0015, Japan}
\affiliation{Graduate School of Science, The University of Tokyo, Bunkyo-ku, Tokyo 113-0033, Japan}

\author[0000-0002-2067-629X]{Takeshi Kamazaki}
\affiliation{National Astronomical Observatory of Japan, National Institutes of Natural Sciences, 2-21-1 Osawa, Mitaka, Tokyo 181-8588, Japan}

\author[0000-0002-7058-7682]{Hideko Nomura}
\affiliation{National Astronomical Observatory of Japan, National Institutes of Natural Sciences, 2-21-1 Osawa, Mitaka, Tokyo 181-8588, Japan}

\author[0000-0002-4124-797X]{Tatsuya Takekoshi}
\affiliation{Kitami Institute of Technology, 165 Koen-cho, Kitami, Hokkaido 090-8507, Japan}

\begin{abstract}

We present the results of data analysis of the [C$\;${\sc i}] ($^3P_1$–$^3P_0$) emission from the $\rho$ Ophiuchi A photon-dominated region (PDR) obtained in the ALMA ACA stand-alone mode with a spatial resolution of 2$\farcs$6 (360~au).
The [C$\;${\sc i}] emission shows filamentary structures with a width of $\sim$1000~au, which are adjacent to the shell structure seen in the 4.5 $\micron$ map.
We found that the 4.5 $\mu$m emission, C$^0$, and CO are distributed in this order from the excitation star (S1) in a complementary pattern.
These results indicate that [C$\;${\sc i}] is emitted from a thin layer in the PDR generated by the excitation star, as predicted in the plane-parallel PDR model.
In addition, extended [C$\;${\sc i}] emission was also detected, which shows nearly uniform integrated intensity over the entire field-of-view (1$\farcm$6$\times$1$\farcm$6).
The line profile of the extended component is different from that of the above shell component.
The column density ratio of C$^0$ to CO in the extended component was $\sim$2, which is significantly higher than those of Galactic massive star-forming regions (0.1--0.2).
These results suggest that [C$\;${\sc i}] is emitted also from the extended gas with a density of $n_{\mathrm{H_2}} \sim 10^3$ cm$^{-3}$, which is not greatly affected by the excitation star.

\end{abstract}


\section{Introduction}

Carbon is one of the major atoms in the Universe.
It is valuable to examine carbon atoms and carbon-bearing molecules to gain deep understanding of the chemistry of the interstellar medium. 
The plane-parallel photon-dominated region (PDR) model (\citealt{Tielens85a, Hollenbach91}) predicts that neutral carbon is formed in a thin layer in the PDR and is transformed to CO in the molecular cloud.
As a result, complementary spatial distributions of [C$\;${\sc i}] and CO emission are expected to appear in a plane-parallel PDR.
Observations, however, show that the spatial distributions of [C$\;${\sc i}] and CO in a Galactic star-forming region are similar (Orion:\citealt{Ikeda02, Shimajiri13}, DR15: \citealt{Oka05}, W51: \citealt{Arikawa99}, Ophiuchi: \citealt{Kamegai03}, RCW38: \citealt{Izumi21}).
The clumpy PDR model (\citealt{Stutzki88, Meixner93}) has been proposed to explain the discrepancy between observations and the plane-parallel PDR model; the model assumes that a molecular cloud has an inhomogeneous density structure and expects that neutral carbon is formed deep inside the molecular cloud.
If the clumpy PDR model is the case, observations with high spatial resolution should be able to identify clumpy structures and reveal differences between the spatial distributions of [C$\;${\sc i}] and CO.
Past observations, in which the spatial resolutions were down to at best 8000~au (0.04~pc; \citealt{Shimajiri13}), of any star-forming regions show similar spatial distributions between the [C$\;${\sc i}] and CO emission in a region. 
No clear evidence has been obtained to conclude which model is correct due to the limitation of the spatial resolution.

One of the best targets to study the PDR is $\rho$ Ophiuchi (Oph) A, illuminated by the Herbig Be star, S1.
It is the closest PDR from the Earth at a distance of $d$=137.3~pc (\citealt{Ortiz-Leon17}), which is closer than Orion by a factor of three ($d\sim$400~pc; \citealt{Menten07}).
\citet{Yamagishi19} observed the $\rho$~Oph~A PDR in $^{12}$CO, $^{13}$CO, and C$^{18}$O ($J$=2--1) lines with ALMA and found clearly layered structures of CO isotopologues, for which CO selective dissociation is considered to be responsible. The finding indicates that the $\rho$~Oph~A PDR has a plane-parallel structure.
As such, $\rho$~Oph~A is an ideal place to study the PDR chemistry because of its simple geometry and closeness, which enables us to perform high spatial-resolution and sensitive observations. 
In this paper, we present our results of the source based on follow-up observations in [C$\;${\sc i}] with ALMA and discuss origins of [C$\;${\sc i}] in the PDR.

\section{Observations and Data Reduction}

We analyzed the ALMA Cycle-6 data at Band 8 for a 1$\farcm$6$\times$1$\farcm$6 region of the $\rho$~Oph~A PDR obtained in the ACA (7-m array + Total Power (TP) array) standalone mode (Project code: \#2018.1.00318.S).
The observed region is the same as that discussed in \citet{Yamagishi19} (RA: 16$^\mathrm{h}$26$^\mathrm{m}$29$\fs$5, Dec: $-$24$\degr$22$\arcmin$43$\farcs$0, J2000) and coincides with a local peak in the low spatial-resolution (2$\farcm$2) [C$\;${\sc i}] map obtained with the Mt. Fuji telescope (\citealt{Kamegai03}).
The observing frequencies were tuned for [C$\;${\sc i}] ($^3P_1$–$^3P_0$) at 492.1607~GHz and for three continuum bands centered at 483.7~GHz (620~$\mu$m) with frequency resolutions of 141.1~kHz and 1.129~MHz, respectively. 
The data reduction was performed, using the Common Astronomy Software Application (\citealt{McMullin07}) version 5.6.1.
In the imaging process of the 7-m array data, the task \texttt{tclean} was employed to recover the extended flux with a spatial grid size of 0$\farcs$39, scales of the \texttt{multi-scale} deconvolver of 0, 5, and 15 pixels (0$\arcsec$, 1$\farcs$95, and 5$\farcs$85), and the \texttt{robust} parameter of 0.
The data of the 7-m array and TP array were combined to recover the total flux, using the task \texttt{feather}.
The final synthesized beam size for [C$\;${\sc i}] was 3$\farcs$32$\times$1$\farcs$90, corresponding to 460~au$\times$270~au for an assumed distance of $d$=137.3~pc (\citealt{Ortiz-Leon17}).
The 1$\sigma$ noise level is 0.45~Jy~beam$^{-1}$ at a velocity resolution of 0.2 km~s$^{-1}$.
We found that continuum, SO$_2$ (7(4,4)--6(3,3)) at 491.93472~GHz, and H$_2$CO (7(1,7)--6(1,6)) at 491.96837~GHz were also detected, in addition to [C$\;${\sc i}], in one of the three ``continuum bands.''
The spatial distributions of these emission are presented in Figure~\ref{othermap}.

\section{Results}

\begin{figure}[ht!]
\centering
\begin{minipage}{0.57\hsize}
\includegraphics[width=\textwidth]{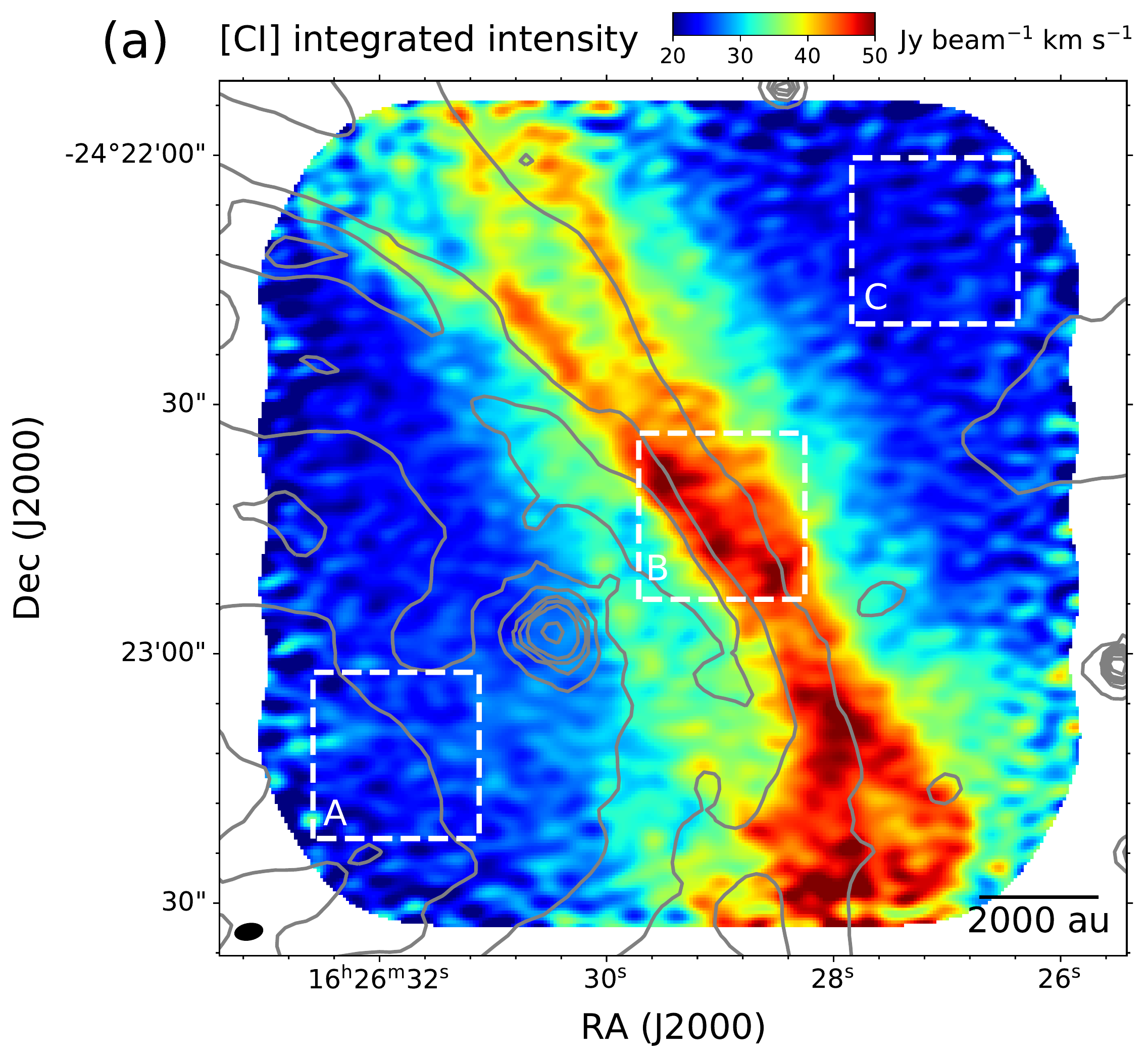}
\end{minipage}
\begin{minipage}{0.41\hsize}
\includegraphics[width=\textwidth]{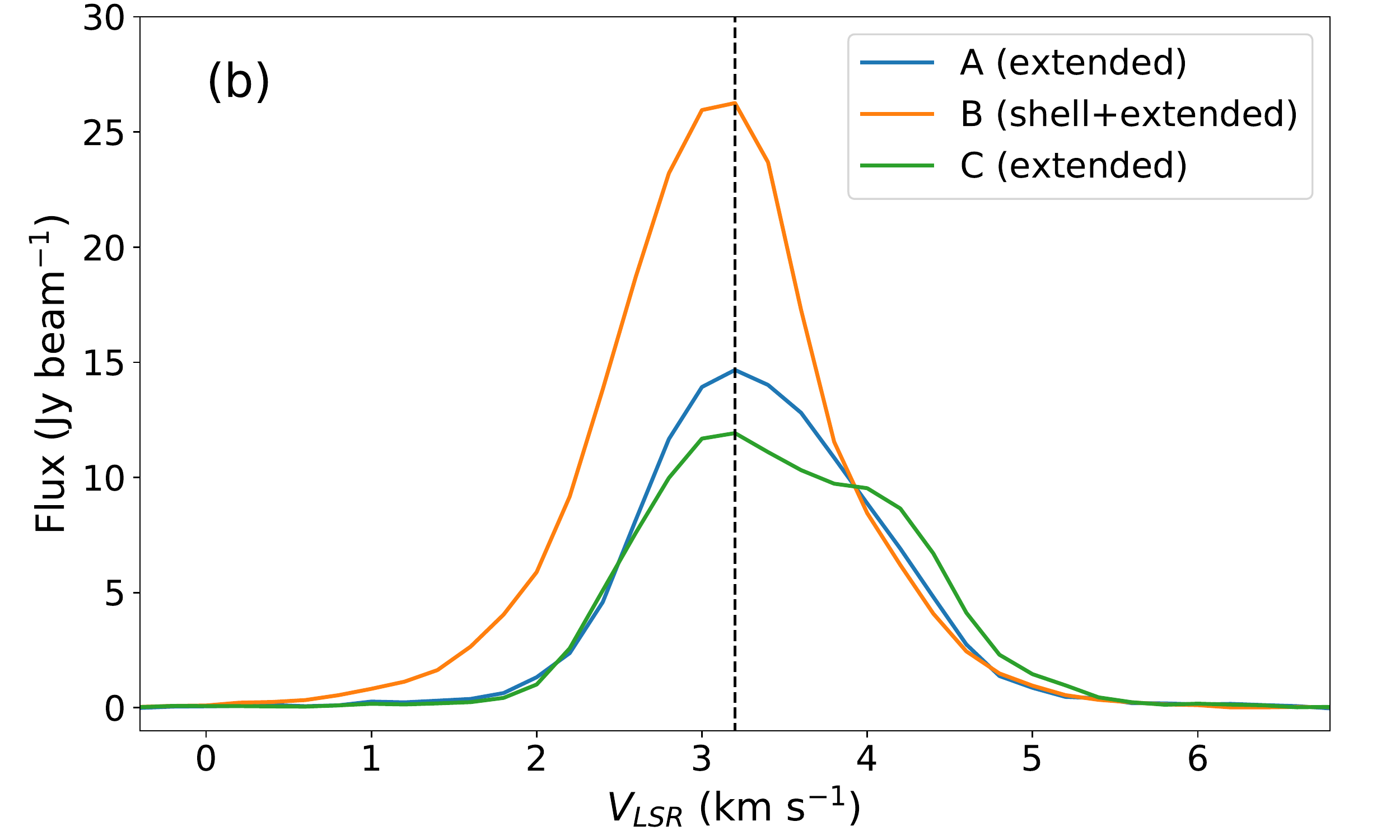}\\
\includegraphics[width=\textwidth]{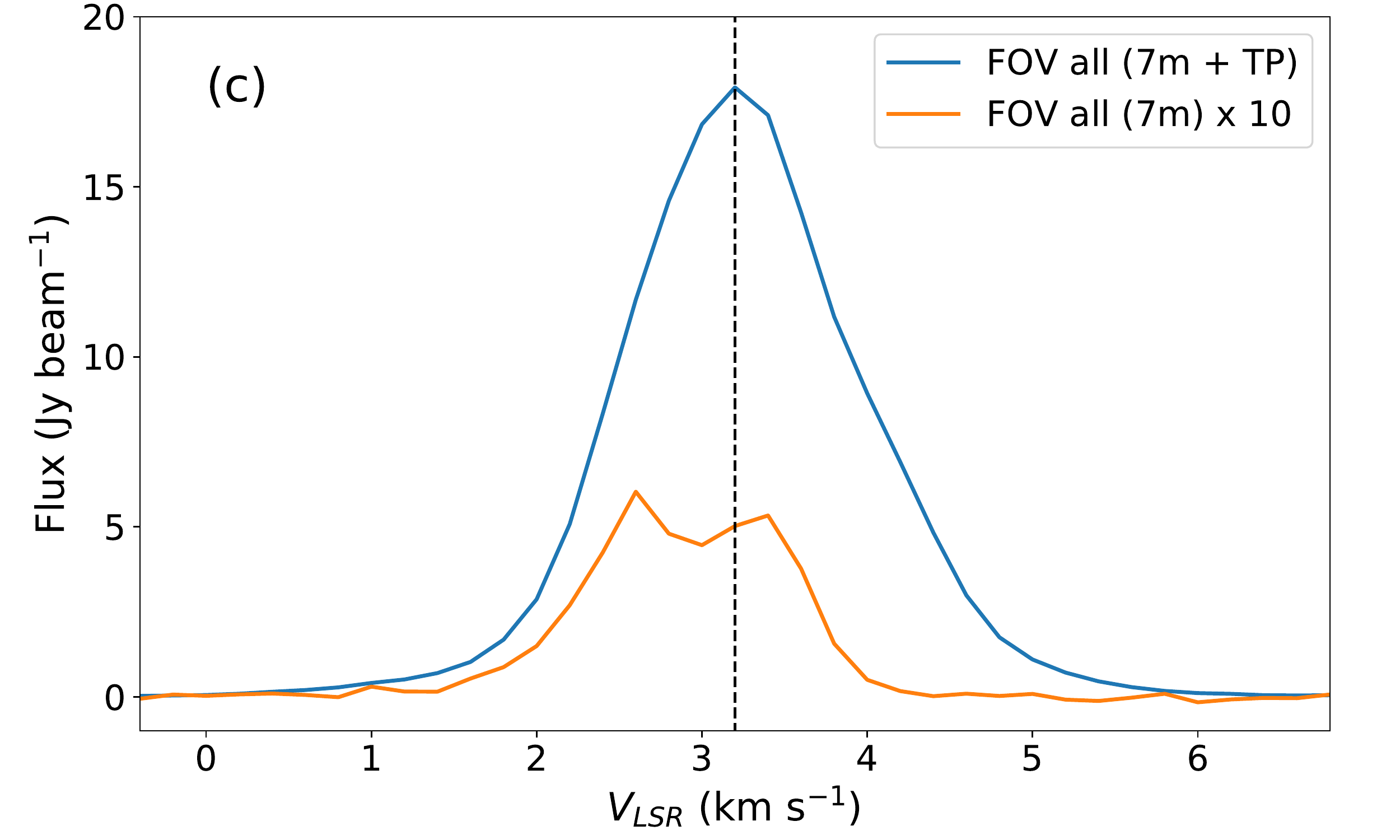}
\end{minipage}

\caption{(a) [C$\;${\sc i}] integrated intensity map over a velocity range of $V_{\mathrm{LSR}}$ = 0.0--6.4 km~s$^{-1}$. Contours indicate Spitzer/IRAC 4.5-$\mu$m intensity (\citealt{Yamagishi19}). The beam size is shown at the bottom left. The excitation star S1 is located outside of this field (RA: 16$^\mathrm{h}$26$^\mathrm{m}$34$\fs$2, Dec: $-$24$\degr$23$\arcmin$28$\farcs$3). (b) Averaged [C$\;${\sc i}] spectra extracted from regions A, B, and C in panel (a). The dashed line indicates $V_{\mathrm{LSR}}$ = 3.2 km~s$^{-1}$ as a reference. (c) Same as panel (b), but those from the entire FOV. Blue and orange lines denote those from the 7m+TP array data and 7-m array data only, respectively. The intensity of the orange one is multiplied by a factor of ten for a display purpose. 
\label{peaksekibun}}
\end{figure}

Figure~\ref{peaksekibun}(a) shows an integrated intensity map of [C$\;${\sc i}] for the $\rho$~Oph~A PDR.
Intense [C$\;${\sc i}] emission is detected from across the entire field-of-view (FOV) with the peak intensity of $\sim$50~Jy~beam$^{-1}$~km~s$^{-1}$.
The integrated intensity map shows filamentary structures distributed in the northeast to southwest direction, which are adjacent to the shell structure seen in the Spitzer/IRAC 4.5-$\mu$m map.
The width of the filamentary structures is $\sim$1000~au, which is smaller by a factor of 8 than the highest spatial resolution in previous [C$\;${\sc i}] observations.
In general, the main emitters in the 4.5-$\mu$m band are Br$\alpha$ and hot dust with a temperature of a few$\times$100~K.
The fact that the narrow emitting region of [C$\;${\sc i}] is distributed adjacent to that of the 4.5-$\mu$m emission indicates that [C$\;${\sc i}] is emitted from a thin layer in the PDR.
The filamentary structures found in the $\rho$~Oph~A PDR are more complicated than the expected structure of a single [C$\;${\sc i}] layer in the plane-parallel PDR model.
The filamentary structures in [C$\;${\sc i}] may reflect filamentary structures of the parental molecular cloud.

We find that the integrated intensities in the faintest part in the FOV (i.e., the northwest and southeast corners in Fig. 1(a)) are similar, $\sim$25~Jy~beam$^{-1}$~km~s$^{-1}$, and that their spatial distributions are almost uniform.
Hence, it is likely that there are two components in the [C$\;${\sc i}] emission in the FOV, namely the filamentary distributed component adjacent to the shell structure (hereafter, shell component) and the uniformly distributed component across the entire FOV (hereafter, extended component).
Considering the intensities and emitting areas of the two components, the extended component is the main [C$\;${\sc i}] source in the FOV.
Indeed, \citet{Kamegai03} detected a highly ($\sim$1 deg) extended [C$\;${\sc i}] emission in $\rho$ Oph.
They explained the extended spatial distribution, using the clumpy PDR model; they assumed that the widely extended emission had some clumpy structures, although they did not resolve any of them, potentially due to the poor spatial resolution of their observations.
In our observation, where the beam size was improved by a factor of 50 from that in \citet{Kamegai03}, no obviously clumpy structures in the northwest and southeast corners were found (Figure~\ref{peaksekibun}(a)).
Therefore, the clumpy PDR model is not suitable to explain the PDR in $\rho$~Oph~A.

Figure~\ref{peaksekibun}(b) shows the spectra extracted from regions A, B, and C, corresponding to (part of) the southeast region in the FOV, shell, and northwest region, which confirm unequivocal detection of the [C$\;${\sc i}] emission.
All the three spectra show a peak velocity of $V_{\mathrm{LSR}}$ $\sim$ 3.2 km~s$^{-1}$, which is comparable with that in C$^{18}$O ($J$=2--1) ($V_{\mathrm{LSR}}$ = 3.0 km~s$^{-1}$; \citealt{Yamagishi19}).
The line profiles are, however, different from region to region; the spectrum from region B, which contains both the shell and extended components, is enhanced in the blue-shifted velocity side ($<$3.2 km~s$^{-1}$), whereas the spectra from regions A and C (extended component only) are enhanced in the red-shifted velocity side ($>$3.2 km~s$^{-1}$). 
The difference implies that the shell and extended components are likely to have different line profiles.
In the direction of $\rho$~Oph~A, \citet{Mookerjea18} identified two structures from self-absorbed [C$\;${\sc ii}] and CO spectra: the main PDR structure at $V_{\mathrm{LSR}}$ = 3.1 km~s$^{-1}$ and a foreground absorber at $V_{\mathrm{LSR}}$ = 3.7 km~s$^{-1}$.
The structures at $V_{\mathrm{LSR}}$ = 3.1 and 3.7 km~s$^{-1}$ might be related with the shell and extended components, respectively.

Figure~\ref{peaksekibun}(c) shows two [C$\;${\sc i}] spectra for the overall FOV.
One is from the 7m+TP array (i.e, flux-recovered) data, the other, from the 7m array data only. 
Since the 7-m array data cannot be used to recover extended structures larger than 14$\arcsec$ (1700~au) in Band 8, the data are expected to be sensitive to the shell component only, whereas the 7m+TP array data are sensitive to both the shell and extended components.
Indeed, the 7-m array data show a much more distinctive enhancement in the blue-shifted velocity side than the 7m+TP array data.
These two spectra reflect differences in the line profiles between the shell and extended components.
Note that the fraction of the flux detected in the 7-m array is only 2~\% of the total (7m+TP array) flux, the fact of which demonstrates importance of the TP array for observations of nearby star-forming regions.

\begin{figure}[ht!]
\centering
\includegraphics[width=\textwidth]{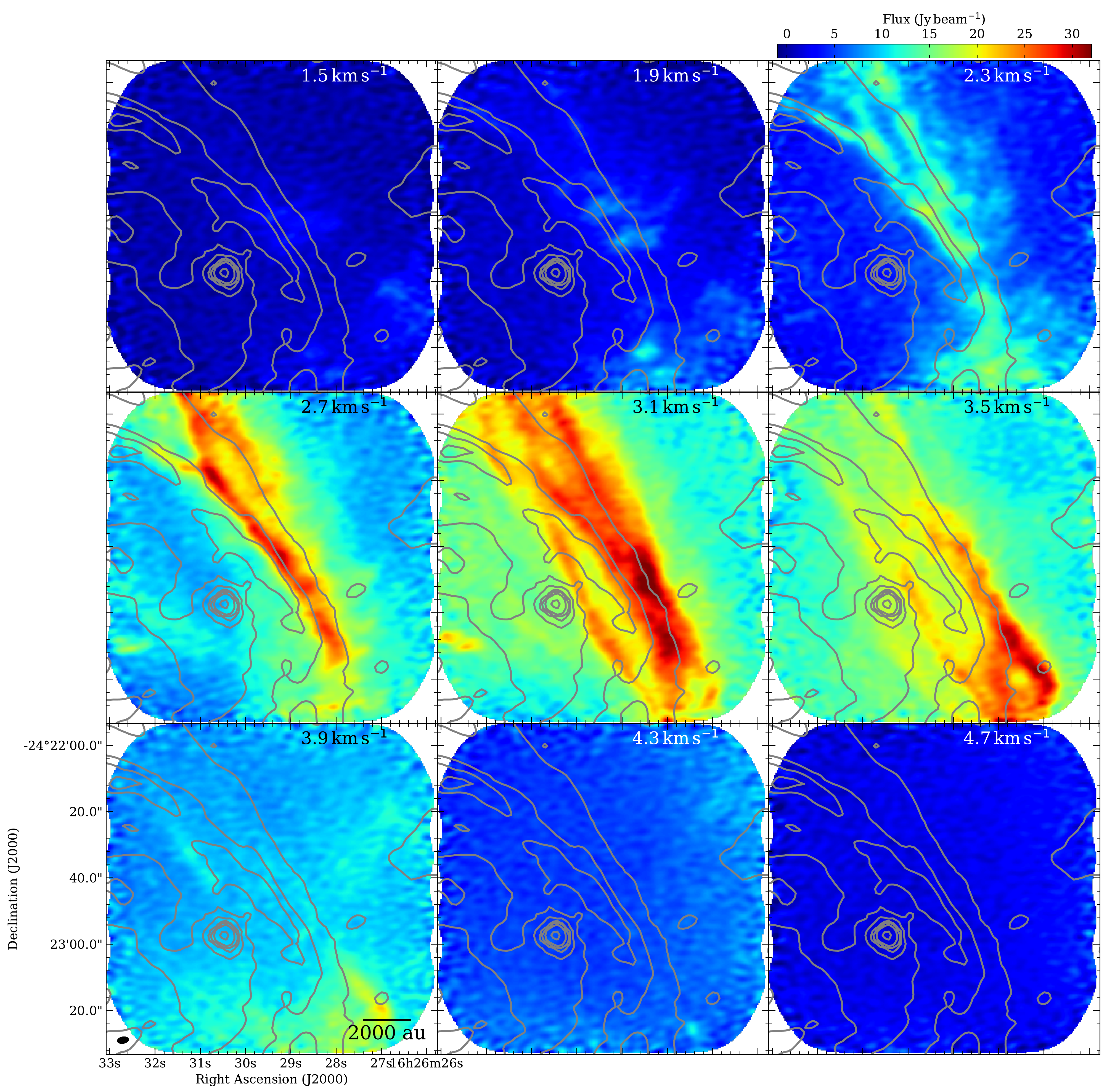}
\caption{Channel maps of [C$\;${\sc i}] in $V_{\mathrm{LSR}}$ = 1.5 -- 4.7 km~s$^{-1}$, where two channels of the original map with a velocity resolution of 0.2 km~s$^{-1}$ are bound up. The central velocity of each channel is indicated at the top right corner. The contours are the same as those in Figure~\ref{peaksekibun}(a). The beam size is shown at the bottom left. \label{chmap}}
\end{figure}

Figure~\ref{chmap} shows channel maps of the [C$\;${\sc i}] emission.
Intense [C$\;${\sc i}] emission is detected with the center velocity of $V_{\mathrm{LSR}}$ = 3.1 km~s$^{-1}$ (see Figure~\ref{peaksekibun}(b)).
In the blue-shifted velocity side, the shell component is stronger than the extended component (e.g, $V_{\mathrm{LSR}}$ = 2.3 km~s$^{-1}$).
Conversely, in the red-shifted velocity side, the extended component is stronger (e.g, $V_{\mathrm{LSR}}$ = 3.9 km~s$^{-1}$).
These channel maps visualize differences in the line profiles seen in Figure~\ref{peaksekibun}(b).

\section{Discussion}

We have identified the shell and extended components in the [C$\;${\sc i}] emission.
In this section, we discuss the origins of the two components, using the column densities of neutral carbon (denoted as\ $N$(C$^0$)) and CO ($N$(CO)), estimated in the following procedures.

The former, $N$(C$^0$), is estimated under an assumption of the local thermodynamic equilibrium (LTE) condition, using formulae in \citet{Ikeda02}.
In the FOV, C$^{18}$O(2-1) was detected in a similar region to [C$\;${\sc i}] (\citealt{Yamagishi19}).
Typically, C$^{18}$O traces low-temperature ($<$100~K) and high-density ($\sim$10$^4$~cm$^{-2}$) environment (e.g, \citealt{Onishi96}).
In such environment, the LTE assumption is expected to be reasonable.
We assume a uniform excitation temperature of $T_{\textrm{ex}}$ = 50~K over the entire FOV.
The estimated value of $T_{\textrm{ex}}$ from ALMA $^{12}$CO($J$=2--1) data was as high as 70~K (\citealt{Yamagishi19}), but the value is not fully reliable due to strong absorption and/or spatial filtering.
\citet{White15} estimated $T_{\textrm{ex}}$ to be 30--50~K from $^{12}$CO($J$=3--2) data with the James Clerk Maxwell Telescope (JCMT).
Although the JCMT data preserve information of the total flux, the spatial resolution is poor (15$\arcsec$).
Our assumed $T_{\textrm{ex}}$ = 50~K is the middle value of these two estimations.
In our estimation, we evaluate the uncertainties of $N$(C$^0$) by varying $T_{\textrm{ex}}$ between 30--70~K.
Consequently, we obtain the optical depth of the [C$\;${\sc i}] emission, $N$(C$^0$), and uncertainty of $N$(C$^0$) to be 0.25-1.2, (3--9)$\times$10$^{17}$ cm$^{-2}$, and 20~\%, respectively.

The other parameter, $N$(CO), is estimated in two complementary methods, using ALMA C$^{18}$O ($J$=2--1) data and Herschel/PACS 70-$\mu$m and 100-$\mu$m data.
Estimation of $N$(CO) from C$^{18}$O ($J$=2--1) is simple, but the total flux is not recovered because TP-array data were not obtained in the observations (Project code: \#2013.1.00839.S).
The indirect estimation of $N$(CO) from the Herschel/PACS data can make use of information of the total flux, but the spatial resolution is comparatively poor (6$\farcs$8 at 100~$\mu$m).
To derive C$^{18}$O-based $N$(CO) ($N$(CO)$_{\mathrm{C18O}}$), the column density of C$^{18}$O ($N$(C$^{18}$O)) is re-estimated using data in the previous study (\citealt{Yamagishi19}) and formulae in \citet{Nishimura15}\footnote{We find that the first factor of equation (10) in \citet{Nishimura15} is larger than the supposed value by two orders of magnitude.}, under an assumption of uniform $T_{\textrm{ex}}$ of 50~K over the FOV.
$N$(CO)$_{\mathrm{C18O}}$ is derived from $N$(C$^{18}$O) using information on the $^{18}$O/$^{16}$O abundance ratio of 490 (\citealt{Wilson94}).
The uncertainty of $N$(CO)$_{\mathrm{C18O}}$ is evaluated in a similar manner as in the estimation of $N$(C$^0$).
In consequence, we derive the optical depth of C$^{18}$O, $N$(CO)$_{\mathrm{C18O}}$, and uncertainty of $N$(CO)$_{\mathrm{C18O}}$ to be 0--0.8, (0--1.2)$\times$10$^{18}$ cm$^{-2}$, and 20~\%, respectively.
The other parameter Herschel-based $N$(CO) ($N$(CO)$_{\mathrm{FIR}}$) is estimated, using Herschel/PACS 70-$\mu$m and 100-$\mu$m data retrieved from the NASA/IPAC Infrared Science Archives (Observation ID: 1342238816) and formulae in \citet{Liseau15}, where a dust emissivity power-law index of 2, dust-to-gas mass ratio of 88 (\citealt{Liseau15}), and $^{12}$CO abundance relative to H$_2$ of 8.3$\times$10$^{-5}$ (\citealt{Frerking82}) are assumed.
Consequently, the $N$(CO)$_{\mathrm{FIR}}$ and its uncertainty are derived to be (0.1--1.4)$\times$10$^{18}$ cm$^{-2}$ and 5~\%, respectively.

\begin{figure}[ht!]
\centering
\includegraphics[width=0.43\textwidth]{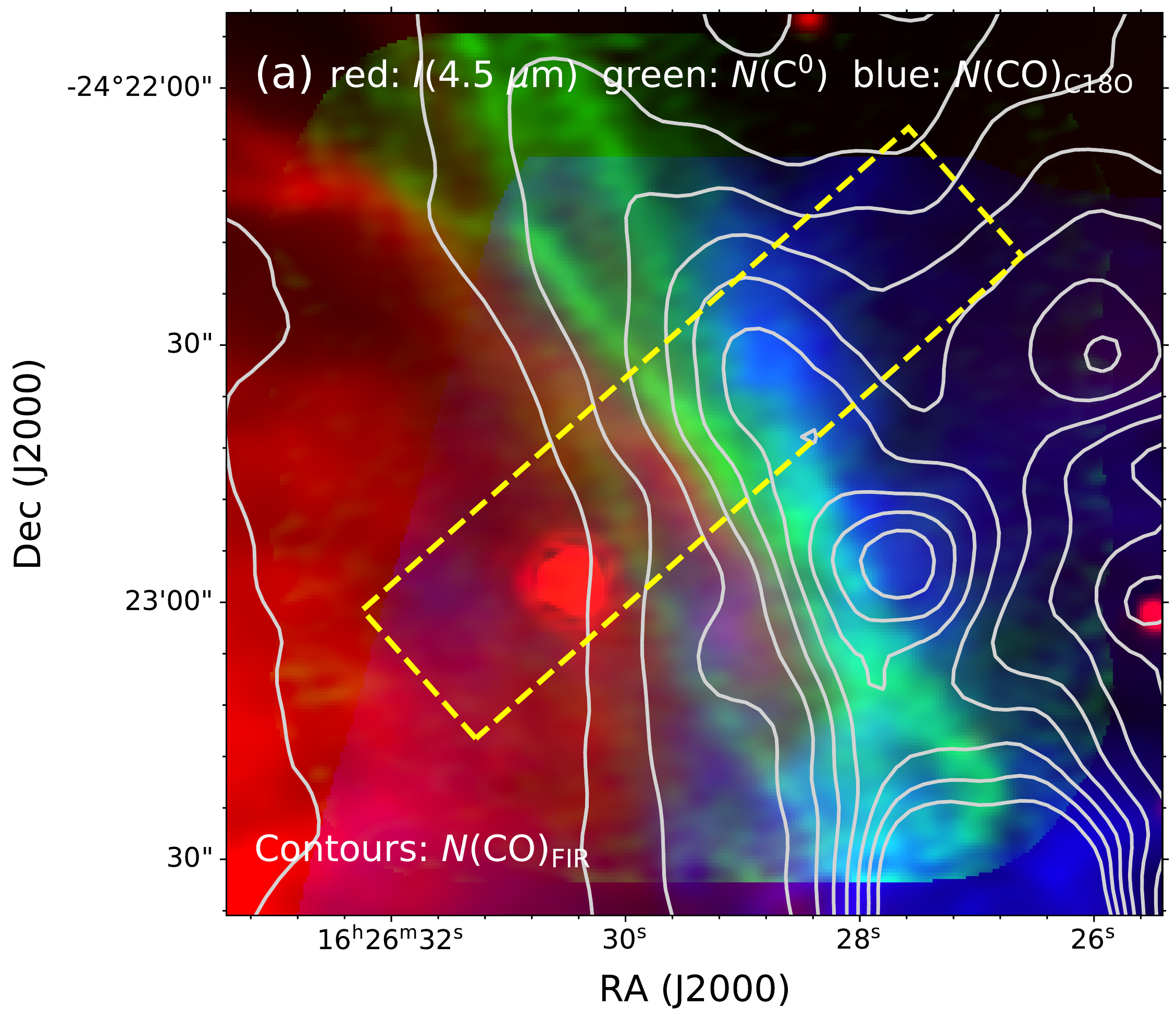}
\includegraphics[width=0.56\textwidth]{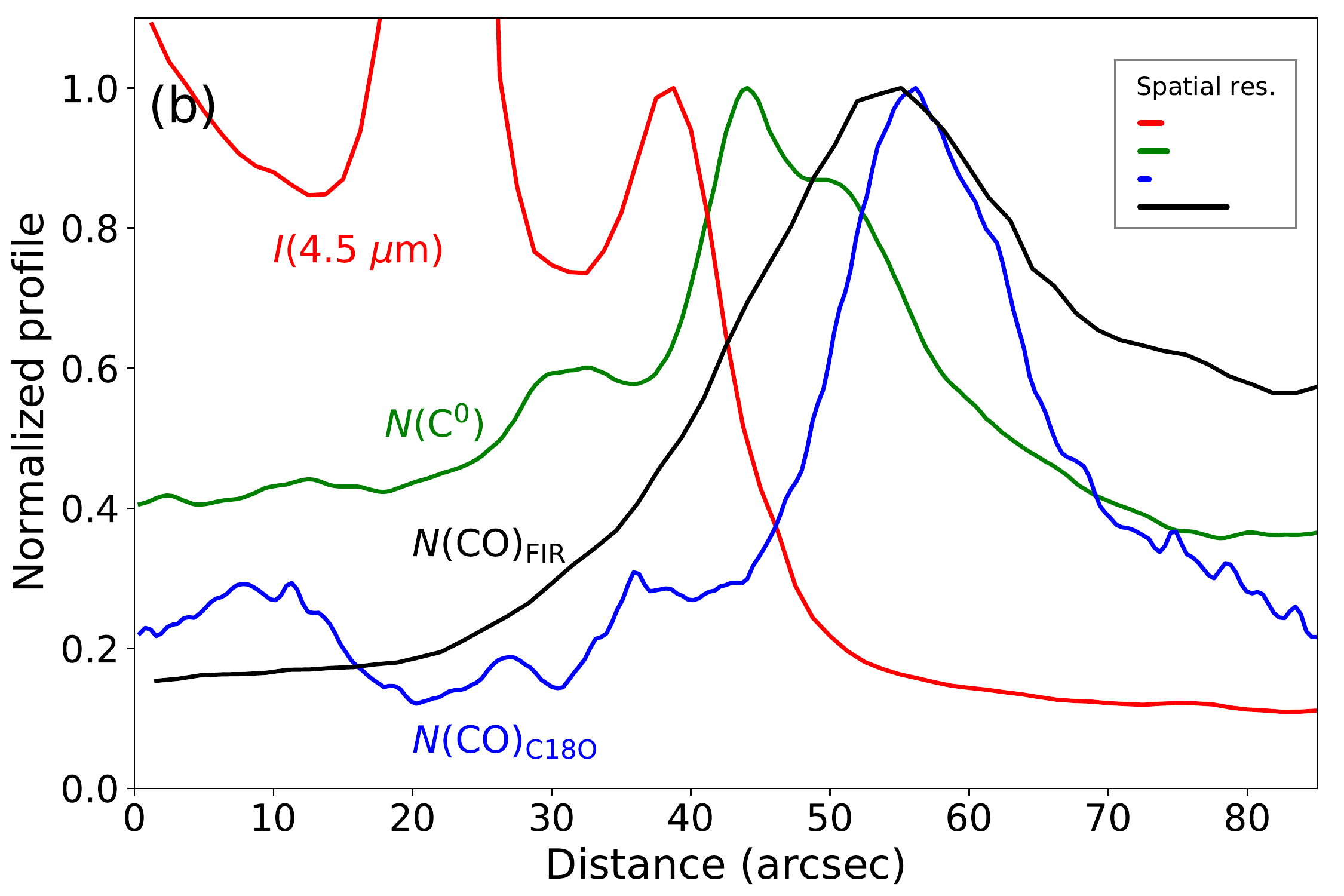}
\caption{(a) Color-composite map of (red) $I$(4.5~$\mu$m), (green) $N$(C$^0$), and (blue) $N$(CO)$_{\mathrm{C18O}}$, where the color ranges are 0--40 MJy~str$^{-1}$, (3--9)$\times$10$^{17}$ cm$^{-2}$, and (0--10)$\times$10$^{17}$ cm$^{-2}$, respectively. Contours indicate $N$(CO)$_{\mathrm{FIR}}$, drawn at linearly-spaced 10 levels in (1--13)$\times$10$^{17}$ cm$^{-2}$. (b) Spatial variations of the average intensity/column densities in the dashed-line rectangular region in panel (a). Profiles are normalized by 24 MJy~sr$^{-1}$ ($I$(4.5~$\mu$m)), 8.0$\times$10$^{17}$ cm$^{-2}$ ($N$(C$^0$)), 9.1$\times$10$^{17}$ cm$^{-2}$ ($N$(CO)$_{\mathrm{C18O}}$), and 9.9$\times$10$^{17}$ cm$^{-2}$ ($N$(CO)$_{\mathrm{FIR}}$). Spatial resolution of each profile is indicated at the top-right\ corner. \label{cutting}}
\end{figure}

Figure~\ref{cutting}(a) shows Spitzer 4.5-$\mu$m intensity ($I$(4.5~$\mu$m)), $N$(C$^0$), $N$(CO)$_{\mathrm{C18O}}$, and $N$(CO)$_{\mathrm{FIR}}$ maps.
We find that $I$(4.5~$\mu$m), $N$(C$^0$), and $N$(CO)$_{\mathrm{C18O}}$ are spatially distributed in this order from the excitation star and in a complementary pattern, the fact of which indicates that [C$\;${\sc i}] is emitted from the PDR generated by the excitation star, as expected from the plane-parallel PDR model.
Both the $N$(CO)$_{\mathrm{C18O}}$ and $N$(CO)$_{\mathrm{FIR}}$ maps show peaks in the northwest side of structures in the $N$(C$^0$) map; this is a confirmation that both the maps show the same CO molecular cloud consistently.
Figure~\ref{cutting}(b) shows one-dimensional spatial profiles of the four maps which are extracted from a region perpendicular to the shell structure (see Figure~\ref{cutting}(a)).
The spatial offsets between the peak positions of $I$(4.5 $\mu$m), $N$(C$^0$), and $N$(CO)$_{\mathrm{C18O}}$ are 700--1400~au (5--10$\arcsec$), indicating that the spatial resolutions in the previous [C$\;${\sc i}] studies were insufficient to distinguish the spatial distribution of [C$\;${\sc i}] from that of CO.
In other words, a combination of [C$\;${\sc i}] and CO data obtained with ALMA with a spatial resolution of a few$\times$100~au, as demonstrated in this work, is essential to study the PDR in detail.

In contrast to the shell component, the extended component is not straightforwardly explained with the plane-parallel PDR illuminated by the excitation star.
If the clumpy PDR model is the case for the extended component as predicted in \citet{Kamegai03}, the extended [C$\;${\sc i}] emission should be spatially resolved into small clumps in observations with high spatial resolution.
However, we have found no such structures in the extended component and the integrated intensity to be uniform.
Since the spatial resolution of our observed data is $\sim$360~au, we cannot fully exclude a possibility that the FOV is filled with much smaller clumps than 360~au.
In such a case, however, UV radiation from the excitation star would penetrate the molecular cloud and produce local fluctuations in the [C$\;${\sc i}] map in a similar way as in the shell component, the deduced view of which is inconsistent with Figure~\ref{peaksekibun}(a).
In addition, the large missing flux in the 7m array (98 \%) indicates that small structures ($<$14$\arcsec$) do not significantly contribute to the [C$\;${\sc i}] emission in the FOV.
Hence, the extended component is likely not to be an ensemble of unresolved small clumpy structures but to be genuinely extended.

One important parameter to discuss the origin of the extended component is the column-density ratio $N$(C$^0$)/$N$(CO) ($R_\mathrm{C/CO}$) because a major formation pathway of C$^0$ is photo-dissociation of CO.
We make a rough estimation of $R_\mathrm{C/CO}$ of the extended component ($R_{\mathrm{ext}}$).
We simply assume that the extended component has uniform $N$(C$^0$) and $N$(CO)$_{\mathrm{FIR}}$ across the entire FOV and that $N$(C$^0$) and $N$(CO)$_{\mathrm{FIR}}$ of the extended component is the minimum value of $N$(C$^0$) (2.9$\times$10$^{17}$~cm$^{-2}$) and $N$(CO)$_{\mathrm{FIR}}$ (1.5$\times$10$^{17}$~cm$^{-2}$), respectively, in Figure~\ref{cutting}(b).
Note that $N$(CO)$_{\mathrm{FIR}}$ is used here because $N$(CO)$_{\mathrm{C18O}}$ is not sensitive to extended emission.
Consequently, $R_{\mathrm{ext}}$ is estimated to be $\sim$2, which is significantly higher than that of the typical $R_\mathrm{C/CO}$ of 0.1--0.2 in the Galactic massive star-forming regions (e.g.,\ Orion: 0.05--0.21, \citealt{Ikeda02}; DR15: 0.02--0.15, \citealt{Oka01}, RCW38: 0.1--0.4, \citealt{Izumi21}).

\citet{Papadopoulos04} showed that the high UV intensity ($G_0$) and low H$_2$ number-density ($n_\mathrm{H_2}$) result in a high $R_\mathrm{C/CO}$.
If the observed high $R_{\mathrm{ext}}$ is caused by a high $G_0$, the $G_0$ value in $\rho$~Oph~A must be higher than those in massive star-forming regions, which is not very likely (cf., $G_0$=10$^5$ in Orion: \citealt{Tielens85b}; $G_0$=(1--4)$\times$10$^3$ in the FOV of the present study: \citealt{Mookerjea18}).
Therefore, the observed high $R_{\mathrm{ext}}$ is likely to be caused by the low $n_\mathrm{H_2}$ gas, which is not greatly affected by the excitation star.
This suggests that the density of the extended component is lower than that in the typical PDR with detection of C$^{18}$O (i.e., $n_{\mathrm{H_2}}<10^4$ cm$^{-3}$).
Hence, the density roughly comparable to the critical density of [C$\;${\sc i}] ($n_{\mathrm{H_2}} \sim 10^3$ cm$^{-3}$; \citealt{Tielens85a}) is likely for the extended component.
In this case, the fact that [C$\;${\sc i}] and CO showed similar spatial distributions in the past [C$\;${\sc i}] studies is plausible, given that CO is a general probe of the molecular gas with a density of $n_{\mathrm{H_2}} \sim 10^{3-4}$ cm$^{-3}$.
As such, [C$\;${\sc i}] is a useful for not only study of PDRs but also for estimating the total gas masses.

\section{Conclusions}

We have analyzed ALMA [C$\;${\sc i}] data obtained for the $\rho$~Oph~A PDR.
We have detected intense [C$\;${\sc i}] emission with a width of $\sim$1000~au, adjacent to the shell structure.
The [C$\;${\sc i}] shell component is located between the 4.5-$\mu$m emission and CO molecular cloud, the fact of which indicates that [C$\;${\sc i}] is emitted from a thin layer of the PDR generated by the excitation star.
The scenario is consistent with the classical plane-parallel PDR model.
We have also detected extended [C$\;${\sc i}] emission with a uniform integrated intensity over the entire FOV.
The extended component has a different line profile from the shell component and a high $N$(C$^0$)/$N$(CO) column density ratio of $\sim$2, suggesting that [C$\;${\sc i}] is emitted from the extended gas with a density of $n_{\mathrm{H_2}} \sim 10^3$ cm$^{-3}$, which is not greatly affected by the excitation star.

\acknowledgments

We express many thanks to the anonymous referee for useful comments.
This paper makes use of the following ALMA data: ADS/JAO.ALMA\#2018.1.00318.S and \#2013.1.00839.S. ALMA is a partnership of ESO (representing its member states), NSF (USA) and NINS (Japan), together with NRC (Canada), MOST and ASIAA (Taiwan), and KASI (Republic of Korea), in cooperation with the Republic of Chile. The Joint ALMA Observatory is operated by ESO, AUI/NRAO and NAOJ.
This work is partially based on archival data obtained with the Spitzer Space Telescope and the Herschel Space Observatory.
Spitzer was operated by the Jet Propulsion Laboratory, California Institute of Technology under a contract with NASA.
Herschel is an ESA space observatory with science instruments provided by the European-led Principal Investigator consortia and with significant participation of NASA.
This work was supported by NAOJ ALMA Scientific Research Grant Numbers 2017-06B.

%




\clearpage
\appendix
\restartappendixnumbering
\section{Other maps}
\label{sec:appendix}

\begin{figure}[ht!]
\centering
\includegraphics[width=\textwidth]{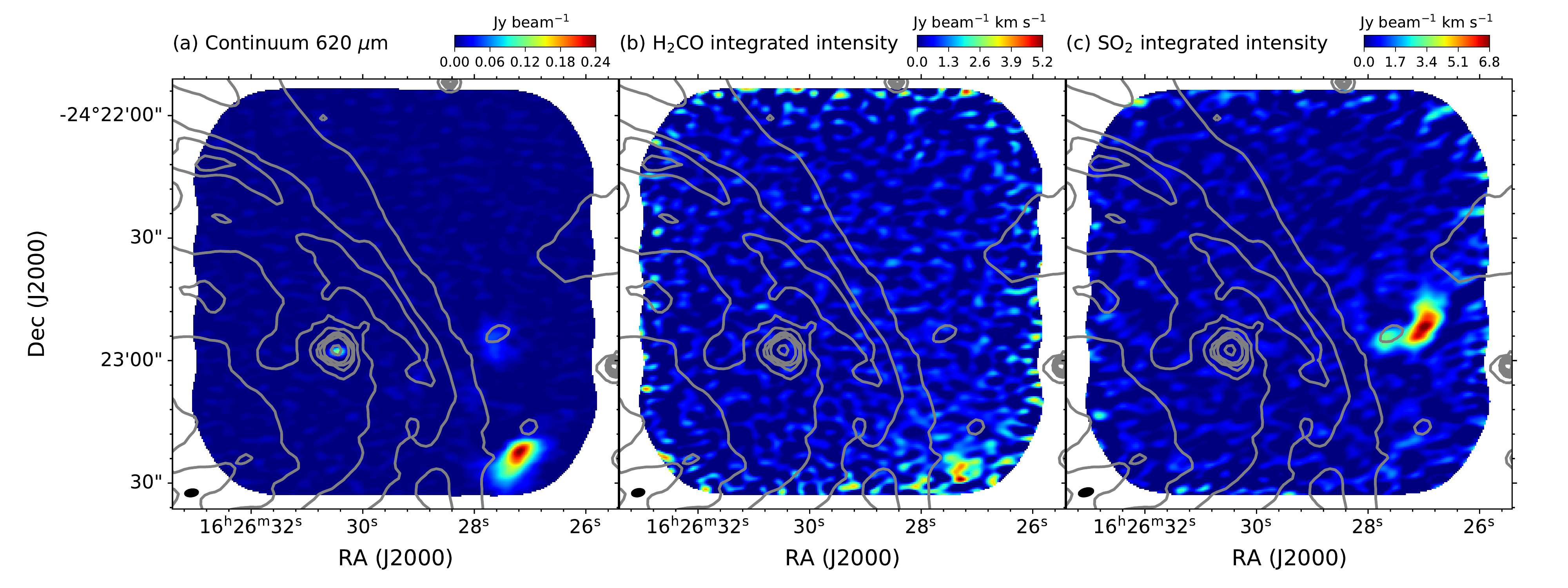}
\caption{Maps of the (a) continuum at 620~$\mu$m and integrated intensities for $V_{\mathrm{LSR}}$ = 2.2--3.8 km~s$^{-1}$ of (b) H$_2$CO and (c) SO$_2$. Contours are the same as those in Figure~\ref{peaksekibun}(a). The beam size is indicated at the bottom-left corner in each panel.
\label{othermap}}
\end{figure}

Figure~\ref{othermap} shows a continuum map at 620 $\mu$m and integrated intensity maps of H$_2$CO and SO$_2$ over a velocity range of $V_{\mathrm{LSR}}$ = 2.2--3.8 km~s$^{-1}$.
The 1-$\sigma$ noise level of the continuum map is 3.9~mJy~beam$^{-1}$, whereas those of the H$_2$CO and SO$_2$ channel maps are both 0.23 Jy~beam$^{-1}$ at a velocity resolution of 0.8~km~s$^{-1}$.
Note that the continuum map is made from the 7m-array data only, while the H$_2$CO and SO$_2$ maps are made from the 7m+TP array data.

\end{document}